\documentclass[12pt]{article}
\usepackage[dvips]{graphicx}
\usepackage{pdproc}

  \makeatletter 
  \def\@cite#1{[#1]} 
  \makeatother    
\begin{document}

\renewcommand{\thefootnote}{\alph{footnote}}

\title{Double Suppression of FCNCs in a Supersymmetric Model
\footnote{Talk given at {\em the 12th International Conference on 
 Supersymmetry and Unification of Fundamental Interactions
 (SUSY 2004),~June~17-23,~2004,~Tsukuba,~Japan}}}

\author{ YUJI KAJIYAMA}

\address{ 
Department of Physics, Kanazawa University \\
Kakuma, Kanazawa, Ishikawa 920-1192, Japan
%%%%% You may comment out the e-mail address line below.  
\\ {\rm E-mail: kajiyama@hep.s.kanazawa-u.ac.jp}}

\abstract{
A concrete model which can suppress FCNCs and CP violating 
phenomena is suggested. It is an $S_3$ symmetric 
extension of the MSSM in extra dimensions, where 
only $SU(2)$ and $SU(3)$ gauge multiplets are assumed to 
propagate in the bulk. 
They are suppressed due to $S_3$ flavor symmetry 
at $M_{SUSY}$, and the infrared attractive force of 
gauge interactions in extra dimensions are used to suppress
them at the compactification scale.  
We find that $O(1)$ disorders of the soft parameters are allowed 
at the cut-off scale to suppress sufficiently 
FCNCs and CP violating phenomena 
\cite{paper1}.
}

\normalsize\baselineskip=15pt

\section{Introduction}

In a Minimal Supersymmetric 
Standard Model (MSSM),~the soft SUSY breaking terms
contain more than 100 
parameters in general. These parameters can induce dangerous 
Flavor 
Changing Neutral Currents (FCNCs) and CP violating phenomena,
which should be
strongly suppressed. 
If these parameters 
approximately satisfy the
conditions 
called universality or alignment, these phenomena are 
suppressed. 
The problem of how and 
why the soft SUSY breaking terms satisfy these 
conditions is called 
the SUSY flavor problem.

There are several theoretical approaches to overcome 
this problem. We will present a model with $S_3$ flavor symmetry 
which is 
embedded into 5 space-time dimensions and show 
that our model can soften the SUSY flavor problem.

\section{$S_3$ Symmetric Model in $D=4$}

We give a brief review of our model at first
\cite{paper1,paper2}.

We consider the $S_3$ flavor symmetric extension 
of MSSM with additional Higgs doublets. 
In our model, there are 3 ``generations" of 
the Higgs doublets and they belong to 
${\bf 2}+ {\bf 1}$ 
representations of $S_3$  which is the same as the matter fields ;
\begin{eqnarray}
{\bf 
2}~~&;&~~Q_I,U_{IR},D_{IR},L_I,E_{IR},N_{IR},
H^U_I,H^D_I 
~~(I=1,2)\nonumber \\
{\bf 1}~~&;&~~Q_3,U_{3R},D_{3R},L_3,E_{3R},
N_{3R},H^U_3,H^D_3 .
\end{eqnarray}
The $S_3$ invariant Yukawa coupling for 
up-quark sector is ;
\begin{eqnarray}
W_U &=&
Y_1^U Q_I H^{U}_3 U_{IR}
+ Y^{U}_{2}f_{IJK} Q_{I}  H^{U}_J  U_{KR}
 + 
Y_3^U Q_3 H^{U}_3 U_{3R}  \nonumber\\
& & + Y^U_{4} Q_3 H^U_I  U_{IR}
+ Y^U_{5} Q_I H^U_I U_{3R}, \nonumber \\  
\mbox{where}~~ f_{121} &=& 
f_{211}=f_{112}=-f_{222}=1~,~
\mbox{others}=0,
%f_{111}=f_{221}=f_{122}=f_{212}=0.
\end{eqnarray} 
and similar for the other 
sectors.

When the electroweak symmetry is broken under the 
assumption
$<H^{U,D}_1>=<H^{U,D}_2>$, the mass matrix for each 
sector becomes
\begin{eqnarray}
{\bf M}_a = \left( \begin{array}{ccc}
m^a_1+m^a_{2} & m^a_{2} & m^a_{5}
\\  m^a_{2} & m^a_1-m^a_{2} &m^a_{5}
    \\ m^a_{4} & m^a_{4}&  m^a_3
\end{array}\right) ~~,~~a=u,d,e,\nu.
\label{mass}
\end{eqnarray}
From the mass 
matrices, we can find a consistent set of mass parameters which 
reproduce the realistic masses and the mixings.

Since both the scalar 
mass terms and the A-terms are $S_3$ invariant, 
the conditions 
universality and alignment are partly realized and 
these have the form  
\begin{eqnarray}
{\bf \tilde{m}^2}_{aLL(RR)} &=&
m^2_{a} \left(
\begin{array}{ccc}
a_{L(R)}^{a} & 0 & 0 \\
0 & a_{L(R)}^{a} & 0 \\
0 & 0 & b_{L(R)}^{a}
\end{array}
\right),\nonumber \\
{\bf \tilde{m}^2}_{aLR} &=&  \left(
\begin{array}{ccc}
m_1^{a} A_1^{a}+m_2^{a} A_2^{a}
 & m_2^{a} A_2^{a} & m_5^{a} A_5^{a} \\
m_2^{a} A_2^{a} & m_1^{a} A_1^{a}-m_2^{a} A_2^{a}
 & m_5^{a} A_5^{a} \\
m_4^{a} A_4^{a} & m_4^{a} A_4^{a} & m_3^{a} A_3^{a}
\end{array}
\right).
\end{eqnarray}    

We can explicitly calculate the parameters $\delta$ 
which are the 
off-diagonal elements of the soft terms in the 
super-CKM basis by using the 
mixing matrices obtained from 
(\ref{mass}). 
The parameters $\delta$ are defined as 
\begin{eqnarray}
\delta_{LL(RR)}^{a} &=&
\frac{U_{aL(R)}^{\dagger} ~{\bf \tilde{m}^2}_{aLL(RR)}~ 
U_{aL(R)}}{m_{\tilde{a}}^2}~\mbox{and}~
\delta_{LR}^{a} =
\frac{U_{aL}^{\dagger}~ {\bf \tilde{m}^2}_{aLR} ~U_{aR}}
{m_{\tilde{a}}^2},
\end{eqnarray}
where ${m_{\tilde{a}}}$ is the average sparticle mass.

We can 
obtain the conditions for the soft parameters
 $\Delta a \equiv b-a$ 
and $\tilde{A}_i-\tilde{A}_j~(\tilde{A^a} \equiv {A^a}/{m_{\tilde{a}}})$ 
by comparing 
these $\delta$'s with the experimental constraints\cite{paper3}
at $M_{SUSY}$. 
The results are 
; 
\begin{eqnarray}
\tilde{A}^\ell_2-\tilde{A}^\ell_2 &\sim& 
O\left(10^{-1}\right)~,~
\mbox{Re}\left( 
\tilde{A}^d_i\right)-\mbox{Re}\left( \tilde{A}^d_j\right)
 \sim 
O\left(10^{-2}\right) ~~~(i,j=1 \sim 5)~,\nonumber \\
\left| 
\mbox{Im}\left( \tilde{A}^u_1\right)\right|
&\sim 
&O\left(10^{-4}\right) ~,~
\left| \mbox{Im}\left( 
\tilde{A}^u_3\right)\right|
\sim O\left(10^{-2}\right) ~, \nonumber \\ 
\left| \mbox{Im}\left( \tilde{A}^d_1\right)\right|
&\sim 
&O\left(10^{-3}\right) ~,~
\left| \mbox{Im}\left( 
\tilde{A}^d_3\right)\right|
\sim O\left(10^{-2}\right) ~, \nonumber \\ 
\Delta a^u_L \Delta a^u_R &\sim &O\left(10^{-1}\right)~,~
\Delta 
a^d_L \Delta a^d_R \sim O\left(10^{-2}\right)~,~
\Delta a^\ell_L \sim 
O\left(10^{-2}\right), \nonumber \\ 
&~&{\mbox{and others}}\sim O(1).
\label{condition}
\end{eqnarray}
It still needs a fine-tuning for the soft parameters 
despite the suppression by $S_3$ symmetry.
\section{Extended Model into $D=5$}

Next, to soften the 
conditions (\ref{condition}) at $M_{SUSY}$ we 
consider a 5 
dimensional model{\cite{paper4}} with orbifold compactification.

\begin{figure}[htb]    
\begin{center}
\includegraphics*[width=8cm]{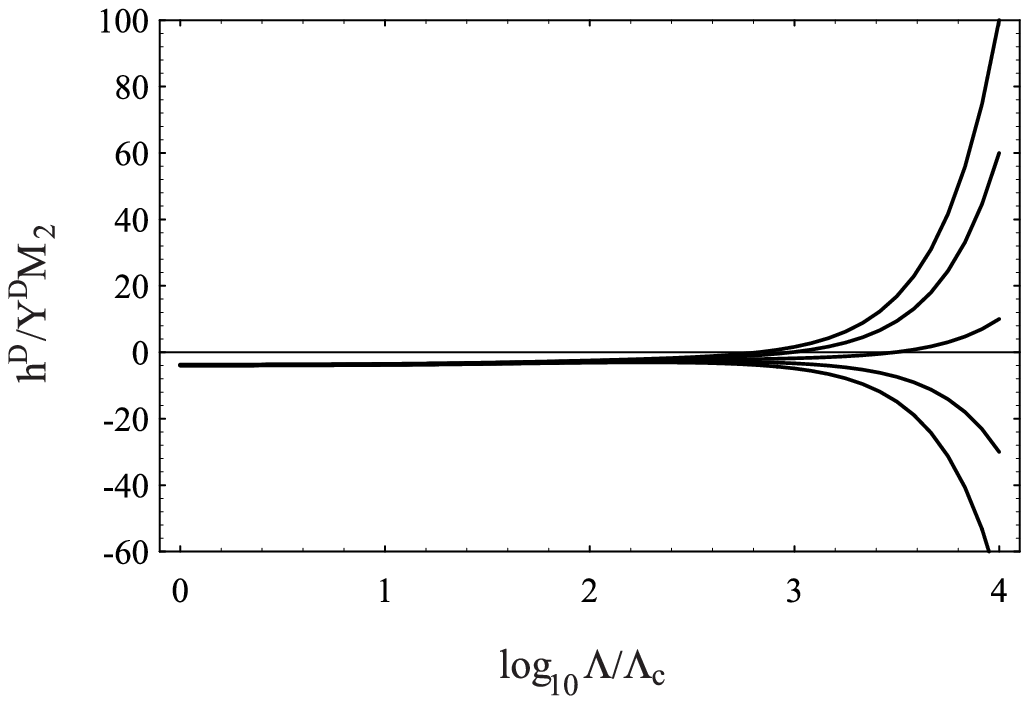}
\caption{%
Infrared convergence of the A-term.
}
\label{fig1}
\end{center}
\begin{center}
\includegraphics*[width=8cm]{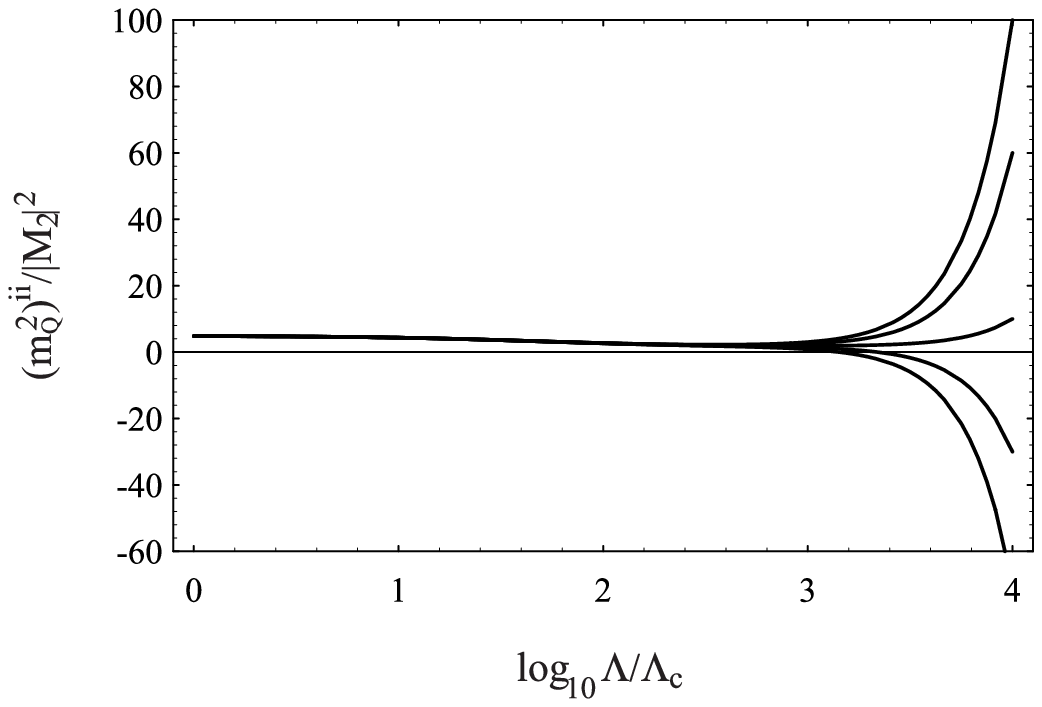}
\caption{%
Infrared convergence of the scalar mass.
}
\label{fig2}
\end{center}
\end{figure}

Let $SU(2)$ and $SU(3)$ gauge fields be the bulk 
fields and matter and 
$U(1)$ gauge fields be the brane fields. 
We assume that the zero modes of the bulk fields consist of 
the previous $4$ dimensional model by some appropriate parity 
assignment. The 
renormalization group 
equations (RGEs) for the ratio of the soft 
parameters and the gaugino mass 
have the infrared fixed points, and 
it converges to the fixed point 
by the power-law running. Moreover 
the convergence is originated by the 
gauge interaction which is 
flavor blind. Therefore the value at low energy scale is 
independent 
of the initial values at the cut-off scale $\Lambda$ and it is 
independent 
of the flavor, too (Fig.~\ref{fig1} and Fig.~\ref{fig2}).   
More precisely, it can be found that the disorder 
of the soft parameters at the compactification scale 
($\Lambda_C \sim$ $M_{SUSY}$) becomes smaller than that at 
the cut-off scale : 
\begin{eqnarray}
\left( \frac{m_{\tilde{a}}^2 ~\Delta a^a}{M_2^2}\right)^{1/2}~,~
\frac{A_i^a-A_j^a}{M_2}~\sim ~O(1)~~{\mbox{at}}~ \Lambda ~
\to
~10^{-2}~~{\mbox{at}}~\Lambda_C ~(\sim M_{SUSY}~\sim ~10^{-4}\Lambda),
\end{eqnarray}    
where $M_2$ denotes $SU(2)$ gaugino mass.

We have to mention the reason why the figures are 
drawn in the range from 
$\Lambda \sim 10^4~\Lambda_C$ to $\Lambda_C$, where $\Lambda_C$ 
denotes the compactification scale of the fifth dimension.
The RGE for the soft mass of the scalar partner of 
the right-handed charged 
lepton (say $m_{E_R}^2$) depends only on the 
brane-localized $U(1)$ gauge field. 
It means that $m_{E_R}^2$ runs logarithmically even in the range 
above $\Lambda_C$ while the gaugino mass decreases in power-law. 
The cut-off scale has been determined by the assumption that 
the ratio ${m_{E_R}^2}/{M_2^2}$ 
satisfies the relation
\begin{eqnarray}
\frac{m_{E_R}^2}{M_2^2}\left(\Lambda_C \right) \sim 10^{-1}~,~ 
\frac{m_{E_R}^2}{M_2^2}\left(\Lambda \right) \sim 10^{+1}.
\label{cutoff}
\end{eqnarray}    
From the condition (\ref{cutoff}), the cut-off scale 
$\Lambda$ can be taken at most to be 
$\Lambda \sim 10^4~\Lambda_C$. 
Because of the power-law convergence of the soft masses, 
the conditions (\ref{condition}) can be softened at $\Lambda$, and 
from the precise estimation, we found that $O(1)$ 
degeneracies of the soft terms 
at $\Lambda$ are necessary to satisfy the condition (\ref{condition}) 
at $M_{SUSY}$.

\section{Conclusion}
We have presented a model which is $S_3$ symmetric 
extension of MSSM. The conditions for the soft terms to suppress FCNCs 
are still severe at $M_{SUSY}$ despite the suppression by 
$S_3$ symmetry. However, at the cut-off scale $\Lambda$, 
$O(1)$ disorder of the soft parameters is allowed, 
because the infrared convergence by power-law running
additionally helps 
realizing universality and alignment of the 
soft parameters. 
%\section{Acknowledgements}

%Acknowledgements should appear just before the references.
%We thank Daisuke Nomura and the authors of Ref.~\cite{paper1} 
%for providing us with a sample equation, figure and table 
%for this template for the SUSY 2004 proceedings.  

\bibliographystyle{plain}

\end{document}